# USING SOCRATIVE TO ENHANCE IN-CLASS STUDENT ENGAGEMENT AND COLLABORATION

Sam M Dakka

Department of Engineering & Math, Sheffield Hallam University, Howard Street, Sheffield S1 1WB,United Kingdom

## ABSTRACT

*Learning and teaching experiment was designed to incorporate SRS-Student Response System to measure and assess student engagement in higher education for level 5 engineering students. The SRS system was based on getting an immediate student feedback to short quizzes lasting 10 to 15 minutes using Socrative software. The structure of the questions was a blend of true/false, multiple choice and short answer questions. The experiment was conducted through semester 2 of yearlong engineering module.  The outcome of the experiment was analyzed quantitatively based on student performance and qualitatively through student questionnaire. The results indicate that using student paced assessments method using Socrative enhanced student's performance. The results showed that 53% of the students improved their performance while 23% neither improved nor underperformed. Qualitative data showed students felt improvement in their learning experience. Overall results indicate positive impact using this technology in teaching and learning for engineering modules in higher education*

## 1.INTRODUCTION

Traditional style lectures due to its passive monotonic one sided delivery rhythm have failed to enthuse and engage students even though the topic might be of great interest. In fact studies showed that the student focus and concentration is totalling meagrely 10 to 15 minutes of one hour session (Bligh, 2000). In addition due to the large fluctuation of registered students each academic year, (Biggs and Tang, 2007) dubbed the traditional lecture as a method of all seasons, most likely due to the advantage by which it can accommodate large number of students. This had prompted the search for more innovative techniques to enhance collaboration and active learning. (Cavanagh, 2011) noted that collaborative learning and active learning raised the bar of interest, stimulated and enriched the student knowledge and kept them attached to the topic under discussion and leveraged their critical thinking skills. (Dewy, 1916) had made the statement "If we teach today as we taught yesterday, we rob our children of tomorrow." interpretation of this statement will lead to the conclusion that we should utilize the current tools available to their disposal in order to stimulate and keep them engaged. Furthermore, (Dahlstrom, 2012), had demonstrated that students advocate strongly for incorporating mobile technologies into their learning environment and consider those tools an integral part of their success and professional accomplishments as students are demanding seamless integration of those technologies. Given the fact that the penetration of these technologies are well spread among student population over variety of devices, such as laptops, smart phones, desktops and tablets, therefore, incorporating mobile technology into learning and teaching might enhance collaborative teaching and enhance active learning stimulating critical thinking skills. Due to the above,  the judgement of binding theory, research and practice is the best way to move forward to achieve the ultimate goal which is stimulate and further enrich students' knowledge through collaborative and active learning and simulating industry environment.

## 2. USING STUDENT RESPONSE SYSTEM-SOCRATIVE

An innovative approach was incorporated into the design and planning of teaching engineering module sessions, Student Response system (SRS) as a part of the lecture. It has been demonstrated that SRS and technology based systems can enhance student engagement (Terrion and Aceti, 2012). SRS has been widely used in the past through devices called clickers which demonstrated positive impact on student engagement (Blasco et al., 2012), were students can give an immediate feedback to the lecturer questions, however clickers has initial cost associated with it in addition it has limited functionality. This has led to search for a tool or software that can be used through smartphones either through mobile connection or Wi-Fi internet connections. The software is called Socrative, and the main advantage of this software lies in its versatility of the type of questions that can be constructed and designed in addition to collaborative features between different teams. The purpose of using Socrative is to assess whether the performance and engagement of the students are enhanced as compared to traditional or other blended teaching and learning lectures. Reflecting on design and planning of teaching session, using Socrative had increased significantly the time of preparation for the session, since construction a large number of variety of questions, such as multiple choice, short answer questions and true/false questions. The element within Socrative that supports student learning and teaching is the fact that the student can get immediate feedback and therefore might reflect further and develop critical thinking approach to various engineering problems.

## 3. LEARNING AND TEACHING, ASSESSMENTS METHODS AND DESIGN OF EXPERIMENTS BASED ON LEARNING THEORIES

During the progress of the module implementation of the constructive alignment theory (Biggs & Tangs, 2007) which transformed my teaching to outcome based learning and teaching was adopted. This is important as the emphasis is on how the student will use the teaching material compared with what contents to incorporate into the module. Also, what learning activities should be implemented to achieve those outcomes and how to assess those activities to measure whether those are attained.

The constructive alignment theory emphasizes the intended learning outcomes of the lecture or module rather what the lecturer is intended to deliver. This is normally written down in a statement called Intended Learning outcomes (ILO's) (Biggs and Tang, 2007) which articulates what the learners will be able to accomplish and the level of accomplishment required (standard). Therefore, reflecting on the delivery sessions based on the constructive alignment theory indicates that one should further develop ideas on how to use Socrative in order to further assess the performance and get the right feedback from the students. Using smartphones which are always on devices (Kolb, 2011), enables immediate feedback and also assessing the students on real time, the ultimate goal using Socrative is to assess how effective incorporating this new technologies in enhancing the in-class and out-class collaboration and consequently the impact on performance and critical thinking skill development. Collaborative learning defined (Prince, 2004) as learning methodology that incorporates meaningful learning with learning engagement process or in other words the students during the learning process reflect about what they are performing and accomplishing, therefore this has an advantage in the sense that the students can control the cognitive process development and also their accomplishments, this definitely a process that stimulates and engage their critical skills which is essential when dealing with complex engineering problems that require innovation and ingenuity. Socrative is a cloud based SRS, versatile in the types of questions administered and the way of implementation can range from individual to groups or teams. It provides immediate feedback for the student and academic, therefore identification of gaps in knowledge and understanding can be addressed almost instantly. (Awdeh et al., 2014) also has shown that students are more engaged and show enthusiasm, so it is this coupled with advanced preparation from the academic that could lead to better performance. Therefore advanced preparation is essential to the success of using this innovative teaching methodology not only in the content but on how to utilize or incorporate the technology effectively during the lecture session, and what approach to implement to get or engage the student to achieve deep learning. Does the software really contribute to that effort of deep learning? This question relates to how student learn, we can answer this question based on teaching theories, two streams of teaching theories phenomenography and constructivism has evolved with time, phenomengraphy (Sonnemann,1954) in the context of learning is based on the student perspective of the learning material rather than what the lecture intended to teach. As compared to constructivism is based on cognitive abilities (Piaget 1950), it implies that learners build and construct knowledge based on their own experience and therefore lectures should be active participatory with other peers including the lecturer. From the above argument related to phenomenography we conclude that we should write down the intended learning outcome so that the student know where they are going and what is the goal on the learning session and see how it contributes to the main project or the topic under consideration. This will definitely transform many students from surface learners (just to get passing grade and move on, or being surface learners due to lack of time) to deep learners. The main challenge at this stage is how to answer this question- how to transform the students to deep learners (Biggs, 1987, Biggs et al., 2001), in general and in particular during session using Socrative. It is worth noting that both theories advocate for learning that promotes conceptual change, not the acquisition of information is the main theme but what we will do with this information. The drive for conceptual change (Biggs and Tang, 2007) can happen if we answer the following questions: (1) Writing down the intended outcomes so student have a clear direction where we are going (2) What is the need to go that path (motivation prerequisite) (3) Students have free ability to concentrate on the task freely (4) collaboration between students, peers and lecturer is essential to develop deep learning. Based on the above 4 questions, drive the transformation from passive to active learners. Elaboration and focus on implementation Socrative in the context of these questions is crucial. In order to answer question 1 we must adhere to strict procedure of writing down a list of verbs from high to low cognitive levels, for example at the top level should be "theorizing" and at the bottom "memorizing" and in between varied levels of cognitive abilities. In deep learning students implement all desired activities both at low and high level but on surface learning they mainly focus on the low level ones, therefore the main task is to try to identify surface learning and make required adjustment to prevent it from happen. Careful examinations of those questions, questions 2 and 3 in the context of the module delivered are clear for the specific subject under consideration as they can identify the progression path on this topic and what is the motivation behind it in addition to further enhancement by a video clip presentation. Furthermore, the students are free to focus on any problems they are encountering. It is question 1 and 4 which are critical for transforming the students into deep learners using Socrative that will be addressed. The quizzes constructed, incorporated as discussed above high and low cognitive verbs that support deep and surface learning, some low level questions require memorizing while others are definitely support deep learning such as verbs related to principles. The questions constructed are supporting variety of levels within the context of deep and surface learning were implemented, it is worth emphasizing question number four with regards to collaboration with peers is planned in the future and will be the subject of future publication.

## 4.RESULTS AND DISCUSSION

### 4.1. Student Performance

Student paced assessments using Socrative was performed for 10 consecutive weeks at semester 2, the entire academic semester for module A. Each assessment lasted between 10 to 15 minutes and was conducted at the same time of the day and at the same day of the week in order to reduce variability associated with human factors that can affect student's performance. The module performance (final grade) of semester two was compared to semester one for each student in order to identify whether using Socrative student paced assessments during semester 2 has improved student performance. Figure 1 illustrates the final grade for each student for semester 1 and 2. The magenta and the blue bars show the student score for semester 1 and 2 respectively. The results of figure 1 demonstrates some improvement in the student performance, in order to further corroborate this observation the data from figure1 was further analysed to account for overall student improvement versus no improvement and less improvement for semester 2 as compared to semester 1, this is illustrated in Figure 2. Inspection of figure 2 reveals that there is 53% improvement in the nominal overall grade performance of the students (first bar, adjacent to vertical axis). Grading both semesters was performed based on the same assessment criteria. The second bar on figure 2 illustrates, under the assumption grading results are within +/-3% precision margin of accuracy, that not only 53% performed better but also significant under performance is just 20%. In overall, Socrative data suggests student paced assessments did improve overall student performance understanding and enhanced student engagement.

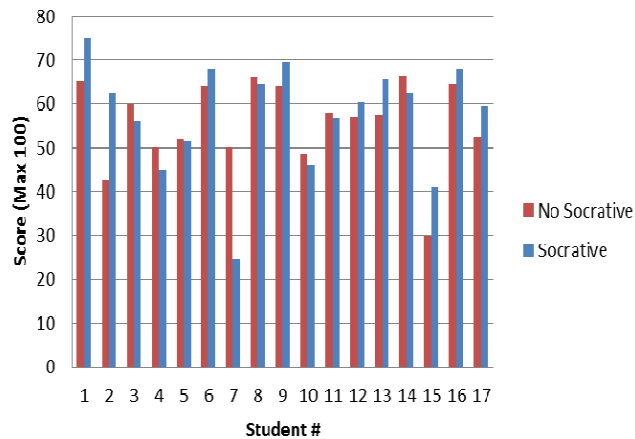

**Figure 1.** Semester 1 & 2 bars respectively for each student number

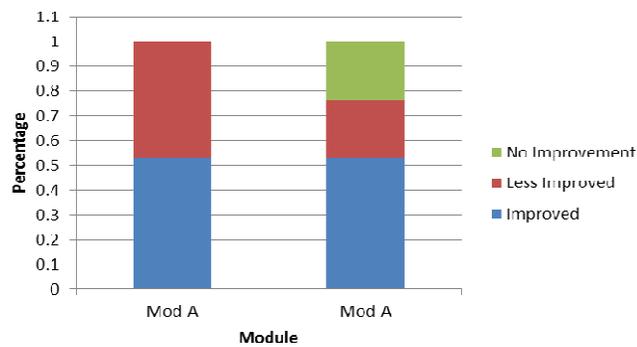

**Figure 2.** Student performance comparison

### 4.2.Student Survey

A student questionnaire was constructed similar to (Awdeh et al., 2014) and posted for students to assess and get feedback on incorporating Socrative in the lecture. It is worth emphasizing that Socrative student paced assessment was conducted during this teaching and learning experiment were the student answered the quizzes privately, to that end there was no any collaboration during the actual experiment with their peers. Despite this lack of live collaboration, table 1 illustrates, there are positive responses and students are embracing and advocate incorporating this technology into their learning and teaching environment. The feedback from the students indicates positively, that the students felt that they are actively collaborating in their learning experience, have the freedom to participate in their learning experience, improved their understanding of material, improved their learning experience, and enhanced the exchange of the information with the lecturer. Answers to questions listed under the question column corresponding to cell number 3,4,7 &8 while did not get the passing mark, this was expected as the design of this learning experiment was not based on collaboration with their peers during the active session.

Table1. Student survey scores based on 100% scale

| Questions/students/average | A | B | C | D | E | F | G | H | Av |
|---|---|---|---|---|---|---|---|---|---|
| I felt that I actively collaborated in my learning experience | 50 | 75 | 75 | 75 | 75 | 75 | 75 | 25 | 66 |
| I felt I have the freedom to participate in my own learning experience | 50 | 50 | 75 | 75 | 75 | 75 | 100 | 50 | 69 |
| In this Method, my classmates and faculty interactions made me feel valuable. | 50 | 50 | 50 | 50 | 0 | 50 | 75 | 25 | 44 |
| This method has favored my personal relationships with my classmates and lecturer | 25 | 25 | 50 | 50 | 25 | 50 | 50 | 25 | 38 |
| Socrative can improve my comprehension of the concepts studied in class | 25 | 25 | 75 | 75 | 75 | 75 | 75 | 75 | 63 |
| The method can lead to a better learning experience | 25 | 75 | 75 | 100 | 75 | 100 | 75 | 75 | 75 |
| Using Socrative gave me the opportunity to have more discussions with classmates | 25 | 50 | 75 | 50 | 0 | 75 | 75 | 25 | 47 |
| Using Socartive allowed the exchange of information with classmates | 50 | 50 | 75 | 25 | 0 | 50 | 75 | 25 | 44 |
| Using Socartive gave me the opportunity of more discussions with the lecturer. | 50 | 50 | 75 | 75 | 0 | 75 | 75 | 25 | 53 |
| Using Socrative allowed the exchange of information with the lecturer. | 25 | 50 | 100 | 50 | 100 | 50 | 100 | 25 | 63 |

The score scale is based on the following respond of the students, strongly agree, agree, neutral, disagree and strongly disagree which granted the marks of 100%, 75%, 50%, 25% and 0% respectively entered in the student survey shown above.

## 5. CONCLUSION

Learning and teaching experiment was designed to incorporate student response system to measure and assess student engagement in higher education for level 5 engineering students. The SRS system was based on getting an immediate student feedback to short quizzes lasting 10 to 15 minutes using Socrative software. The structure of the questions was a blend of true/false, multiple choice and short answer questions. The experiment was conducted through semester 2 of yearlong module. The outcome of the experiment was analyzed quantitatively by comparing the overall performance of the students on semester 2 as compared to 1 and qualitatively through student questionnaire. The results indicate that using this method (student paced assessments using Socrative) despite the lack of collaboration aspects designed into the experiment did enhance the performance of the students on semester 2 as compared to semester 1. The results showed that 53% of the students improved their performance while 23% neither improved nor underperformed as compared to semester 1. So a total 76% improvement. Qualitative data based on the student questionnaire showed that the student felt that this method improved their learning experience, actively collaborated and have the freedom to participate in their learning experience and allowed the exchange of information with the lecturer. The method of team work, student paced or lecturer paced was not implemented in the teaching experiment yet, therefore the student expressed less positive responses to the elements of interaction, relationship and exchanging information with their class mates which was expected and will be the subject of future research and publication. Future work will be focused on building upon team work configuration within Socrative in the context of addressing all the elements of deep learning transformation discussed. Overall the results indicate positive impact of using this technology in teaching and learning for engineering modules in higher education.

## ACKNOWLEDGEMENTS

The authors would like to thank students on the engineering module who participated in this teaching and learning experiment.